\documentclass[journal,10pt]{IEEEtran}
\usepackage[utf8]{inputenc}
\usepackage{cite}
\usepackage{adjustbox}
\usepackage{amsmath,amssymb,amsfonts}
\usepackage{algorithmic}
\usepackage{graphicx}
\usepackage{tikz}
\usepackage{scalefnt}
\usepackage{fix-cm} 
\usepackage{multirow}
\usepackage{rotating}
\usepackage{textcomp}
\usepackage{mathrsfs}
\usepackage{xcolor}
\usepackage{blindtext}
\usepackage{pgfplots}
\usepgfplotslibrary{groupplots}
\usepackage{cleveref}
\usepackage{makecell}
\usepackage{authblk}
\usepackage{verbatim}
\usepackage{algorithm}
\usepackage{algorithmic}
\usetikzlibrary{arrows.meta}

\newcommand\raisepunct[1]{\,\mathpunct{\raisebox{0.3ex}{#1}}}
\pgfplotsset{compat=1.14}
\begin{document}
  \title{Deep unfolding of the weighted MMSE beamforming algorithm }
  \author{
    \IEEEauthorblockN{Lissy Pellaco, Mats Bengtsson, and Joakim Jaldén
    \thanks{This work was supported by European Research Council (ERC) under the European Union's Horizon 2020 research and innovation programme (grant agreement No 742648)}
    \thanks{
    Lissy Pellaco, Mats Bengtsson and Joakim Jald\'en are with the Division of Information
    Science and Engineering, School of Electrical Engineering and Computer Science, KTH Royal Institute of 
    Technology, Stockholm 11428, Sweden (e-mail: pellaco@kth.se, matben@kth.se, and jalden@kth.se).}
    }}

\maketitle
\begin{abstract}
Downlink beamforming is a key technology for cellular networks. However, computing the transmit beamformer that maximizes the weighted sum rate subject to a power constraint is an NP-hard problem. As a result, iterative algorithms that converge to a local optimum are used in practice. Among them, the weighted minimum mean square error (WMMSE) algorithm has gained popularity, but its computational complexity and consequent latency has motivated the need for lower-complexity approximations at the expense of performance. 

Motivated by the recent success of deep unfolding in the trade-off between complexity and performance, we propose the novel application of deep unfolding to the WMMSE algorithm for a MISO downlink channel. The main idea consists of mapping a fixed number of iterations of the WMMSE algorithm into trainable neural network layers, whose architecture reflects the structure of the original algorithm. With respect to traditional end-to-end learning, deep unfolding naturally incorporates expert knowledge, with the benefits of immediate and well-grounded architecture selection, fewer trainable parameters, and better explainability. 
However, the formulation of the WMMSE algorithm, as described in Shi et al., is not amenable to be unfolded due to a matrix inversion, an eigendecomposition, and a bisection search performed at each iteration. Therefore, we present an alternative formulation that circumvents these operations by resorting to projected gradient descent. By means of simulations, we show that, in most of the settings, the unfolded WMMSE outperforms or performs equally to the WMMSE for a fixed number of iterations, with the advantage of a lower computational load.

\end{abstract}

\section{Introduction}
Downlink beamforming is a pivotal technology in fourth and fifth generation cellular communication systems~\cite{5G,Ericsson}.
It leverages the use of multiple antennas to achieve an improved spectral efficiency to meet the demanding performance requirements expected from the system~\cite{ericsson20175g}. 
A common approach to downlink beamforming design is to find a transmit beamformer that maximizes the weighted sum rate (WSR) under a total transmit power constraint. However, the WSR maximization problem subject to a power constraint is known to be NP-hard~\cite{Liu2011,Luo2008}.
Algorithms that find the optimal solution exist~\cite{Joshi2012,Liu2012,Bjornson2012}, but the high computational complexity and the consequent latency and power consumption that they exhibit, especially as the number of users and antennas grows, negate the advantages of beamforming. 
Therefore, it is common to resort to suboptimal solutions that mitigate the computational complexity at the expense of performance.
On one hand, simple heuristics with closed form solutions have been proposed, such as zero forcing beamforming and its regularized form~\cite{Peel2005}, maximum ratio transmission~\cite{Lo1999}, matched filtering and Wiener filtering precoders~\cite{Joham}. These heuristics reduce the computational load, the execution time, and the power consumption, but limit the achieved performance gain.
On the other hand, there exist iterative algorithms, based on convex approximations~\cite{Ng2010,Tran2012,Kibria2013} or on tractable alternative formulations~\cite{ Shi2011,Christensen2008,Schmidt2009} of the original problem, that converge to a local optimum. They are more onerous in terms of computational load and power requirements and exhibit a longer delay, but achieve a higher WSR. A popular approach that belongs to this class of algorithms is the iterative weighted minimum mean square error (WMMSE) algorithm.
It converges to a local optimum of the WSR function and, as a consequence of the resulting performance, it has gained popularity. However, it exhibits relatively high computational complexity. This has fostered the development of lower-complexity approximations at the cost of a degraded performance~\cite{Nguyen2014a,Carvalho2012}.
In fact, the computational complexity versus performance trade-off is of considerable importance for cellular networks. Base stations must comply with stringent cost and power specifications and must, at the same time, respond quickly to the fast changing  channel conditions. Moreover, with the fifth generation cellular networks, they must accommodate latency requirements of newly supported real-time applications, from industrial automation to remote medical care~\cite{5g-internet}. 

In this paper, we address this computational complexity versus performance trade-off by applying a machine-learning-based technique, called \textit{deep unfolding}, to the WMMSE algorithm described in\cite{Shi2011} for a multiple-input single-output (MISO) downlink channel.

\subsection{Relevant Prior Work}

Inspired by the recent advances in machine learning and by its promising results in physical layer applications~\cite{Qin2019,Dorner2017,Zhang2019survey,Gunduz2019,Hoydis2017,Chen2019,Zappone2019}, machine-learning-based solutions that address the complexity versus performance trade-off have been proposed for downlink beamforming~\cite{Huang2018,Zhang2019b,Lin2020} and, in particular, for the WMMSE algorithm~\cite{Sun2018,Xia_2019,Huang2019}. 
The common underlying idea consists of replacing the well-performing, yet expensive and high-latency, iterative algorithms with neural networks. These approaches are based on end-to-end learning, i.e., neural networks take as input the wireless channel and directly predict the beamformer weights. In this case, the complexity and latency constraints translate into architectural constraints. In fact, the computational load and the latency of the  inference process are predetermined by the network architecture. Therefore, the network can be designed to be compliant with the power consumption and real-time application requirements at the expense of bounding its performance. 

With regards to the WMMSE algorithm, neural networks have been employed both as i) fixed-complexity function approximators and as ii) comparably performing alternatives within the complexity constraints.
In~\cite{Sun2018}, the authors train a fully connected deep neural network to approximate the output of the WMMSE algorithm and provide theoretical conditions under which iterative optimization algorithms can be approximated by fully connected deep neural networks. In~\cite{Xia_2019}, the authors use the WMMSE algorithm to pre-train the weights of a neural network that predicts the transmit beamformer of a MISO channel. In the pre-training phase, the performance of the network is clearly bounded by the solution obtained by the WMMSE algorithm. Therefore, the pre-training is followed by direct maximization of the WSR function with the purpose of approaching the global optimum. In~\cite{Huang2019}, the authors design a deep neural network to find the beamformer weights and evaluate its performance in case of i) supervised learning, i.e., the network is trained to approximate the WMMSE solution, and ii) unsupervised learning, i.e., the network is directly trained by maximizing the WSR function. 

It must be noted that in case of supervised learning the performance of the neural network is bounded by construction by the WMMSE performance. Conversely, in case of unsupervised learning the performance is not restricted and the network could in principle achieve an improved WSR. 
However, neither~\cite{Xia_2019} nor~\cite{Huang2019} manages to surpass the WSR attained with the WMMSE algorithm.
Furthermore, these approaches share the problem of selecting the proper neural network architecture. Although the computational constraints translate into architectural constraints, the search space remains extremely large. It is well known that the architecture has a large impact on the generalization capability and thus the performance of a neural network. 
Selecting the proper architecture is still an open issue and considerable research efforts are devoted to this~\cite{Suganuma2017,Hunter2012,Benardos2007,Luo2018,Sun2018}. Meanwhile, it is normal practice to rely on intuition, experience, and the trial and error approach. 
Another drawback that is common to~\cite{Xia_2019,Huang2019,Sun2018} is related to the explainability of the neural network behavior as the proposed approaches are based on end-to-end learning. Albeit the rationale behind the training process is clear -- whether the training is supervised, unsupervised, or a combination of both --  the resulting behavior is not yet explainable, and thus the network is regarded as a black-box~\cite{Samek2017}. In the context of wireless communications, this lack of explainability hinders the deployment of deep-learning-based solutions. Communication engineers usually rely on  expert knowledge developed from a solid theoretical background, measurements, past experience, and a deep understanding of the system. Therefore, the introduction of end-to-end learning solutions, especially for critical applications, such as autonomous driving, encounters skepticism in the wireless networks community~\cite{Shafin2019}.
It must be mentioned that in~\cite{Xia_2019} the authors, instead of directly predicting the beamformer vector, make use of expert knowledge to predict only a set of features from which the beamformer vector can be constructed. However, the lack of explainability of the network behavior -- whether the network predicts the entire beamformer vector or a subset of features thereof -- remains.
\subsection{Deep Unfolding}
Recently, an alternative learning approach has enabled significant advancements in the trade-off between complexity and performance for iterative algorithms.
The fundamental idea consists of i) mapping the iterations of an optimization algorithm into trainable neural network layers whose architecture follows the structure of the original algorithm, ii) fixing the number of iterations, i.e., the number of layers, according to the computational complexity and latency requirements, and iii) optimizing the trainable parameters of the network in order to achieve the best possible performance within the fixed number of iterations.
This approach takes the name of \textit{deep unfolding} and was pioneered by Gregor and LeCun in 2010~\cite{Gregor2010}. Since then, this same idea has been successfully applied in various fields, e.g., sparse coding~\cite{Gregor2010,Sprechmann2012,Pla2018}, sparse signal recovery~\cite{Borgerding2017,Ito2019,Yao2019,Khobahi2019}, image restoration~\cite{Bertocchi2020}, MIMO detection~\cite{Wiesel2019,Un2019,Takabe2018,Khani2019,He2020}, decoding~\cite{Wadayama2019,Wei2020,He_turbonet,Mengke2019}, compressed sensing~\cite{Zhang_2018_CVPR,yangNIPS2016}, and 
1-bit precoding~\cite{Stimming2019}.

Deep unfolding allows for naturally incorporating expert knowledge in the learning approach~\cite{Hershey2014}. This has multiple  benefits. 
First, by fixing the neural network structure, deep unfolding allows researchers to drastically reduce the time dedicated to test different architectures and provides them with a strong motivation behind the choice of the adopted architecture. Clearly, the challenge of selecting the trainable parameters within the fixed network structure remains, but the choice, in most of the cases, is naturally guided by knowledge of the underlying algorithm.
Second, the number of trainable parameters is significantly reduced with respect to end-to-end learning approaches, as the structure of the algorithm remains intact and a usually small subset of parameters is selected to be trainable.
Third, in areas rich of expert knowledge, such as wireless communications, the introduction of deep-unfolding-based solutions is widely favored with respect to end-to-end learning solutions. 
In fact, in this case, the structure of algorithms  well-established in the field  is maintained and is reflected in the structure of the network, resulting in a higher degree of explainabilty~\cite{Monga2019}.
Finally, the performance of the unfolded neural network might benefit from the same performance guarantees that hold for the original algorithm~\cite{Balatsouka2019}.

\subsection{Contributions}
Our main contribution is the novel application of deep unfolding to the WMMSE algorithm in the case of a MISO downlink channel, in order to address the complexity versus performance trade-off.
 As noted in~\cite{Sun2018}, deep unfolding cannot be directly applied because the WMMSE algorithm as described in~\cite{Shi2011} includes complex operations, such as matrix inversions, eigendecompositions, and bisection searches. As a result, the authors of~\cite{Sun2018} choose to adopt the end-to-end learning approach over deep unfolding and use a fully connected neural network to approximate the WMMSE algorithm. Conversely, we propose a formulation of the WMMSE algorithm that involves only vector multiplications, summations, and easy-to-implement projections, and thus allows for deep unfolding. We circumvent matrix inversions, eigendecompositions, and bisection searches by replacing the method of Lagrange multipliers, applied in the original WMMSE algorithm, with the projected gradient descent (PGD) approach~\cite{Bertsekas99}. As a result, our formulation of the WMMSE algorithm i) can be easily unfolded and optimized within the computational and latency constraints, ii) replaces the heavy operations in the original WMMSE algorithm with simple vector operations, and iii) can be efficiently implemented on standardized hardware platforms specialized for deep learning processing, which are expected to replace expensive and dedicated integrated circuits in base stations.
In order to meet the power consumption and latency requirements at the base station, we fix the number of PGD steps per WMMSE iteration and the total number of iterations accordingly. Clearly, this comes with a degradation of performance. In order to compensate for it and achieve the best possible performance within the computational constrains, we select the PGD step sizes as trainable parameters. 
By means of extensive simulation results, we show that the step sizes can be effectively learnt and that, in most of the settings, the unfolded WMMSE algorithm outperforms or performs equally to the original WMMSE algorithm for a given number of iterations, with the additional benefits of lower computational complexity and feasible implementation on standardized hardware.

To encourage reproducibility, the code to generate the results in the paper is available at~\cite{GITHUB}.

\subsection{Notation}
We adopt the following notation in this paper. We indicate matrices by bold uppercase letters and column vectors by bold lowercase letters. $\boldsymbol{0}$ is a column vector of zeros and $\boldsymbol{I}_M$ is the $M\times M$ identity matrix. We use $\boldsymbol{I}$ when the size of the matrix can be understood by context. We use $\left\lVert \cdot \right\rVert$ to indicate the Frobenius norm of a matrix. We use the superscripts $(\cdot)^T$, $(\cdot)^H$, and $(\cdot)^{-1}$ to indicate the transpose, the Hermitian transpose, and the inverse of a matrix, respectively.  $\mathtt{Tr}(\cdot)$ indicates the trace of a matrix and $\mathtt{diag}(\cdot)$ indicates the diagonal elements of a matrix. $\mathbb{R}^{M}$ indicates the $M$-dimensional real space and $\mathbb{C}^{M}$ indicates the $M$-dimensional complex space. $\Re(\cdot)$ and $\Im(\cdot)$ denote the real part and the imaginary part, respectively. $\mathbb{E}_y(x)$ indicates the expected value of $x$ computed with respect to the probability distribution of $y$. $\nabla f$ is the gradient of $f$. The complex Gaussian distribution with mean $\mu$ and variance $\sigma^2$ is denoted by $\mathcal{CN}(\mu,\,\sigma^2)$ and analogous notation follows for the multivariate complex Gaussian distribution. 

\section{Problem formulation and system model}
We consider a multiple-input single-output (MISO) interference downlink channel. The base station, equipped with $M$ transmit antennas, sends independent data symbols to $N$ single-antenna users. We indicate vectors and matrices relative to the $i^{th}$ user with index $i$.
Let $x_i \sim \mathcal{CN}(0,\,1)$ be the transmitted data symbol and let  $\boldsymbol{h}_i \sim \mathcal{CN}(\boldsymbol{0},\,\boldsymbol{I}_M)$  be the channel.
With linear beamforming, the signal at the receiver is 
\begin{equation}
y_{i} = {\boldsymbol{h}^{H}_i}\boldsymbol{v}_{i}x_{i} + \sum_{j=1,j \neq i}^{N}{\boldsymbol{h}^{H}_i}{\boldsymbol{v}_{j}x_{j}} + n_{i},
\end{equation}
where $\boldsymbol{v}_i \in \mathbb{C}^{M}$ is the transmit beamformer for user $i$ and where $n_i \sim \mathcal{CN}(0,\,\sigma^{2})$ is independent additive white Gaussian noise with power $\sigma^2$. The signal-to-interference-plus-noise-ratio (SINR) is 
\begin{equation}\label{eq:SINR}
\mathrm{SINR}_i = \frac{|\boldsymbol{h}^H_{i}\boldsymbol{v}_i|^2}{ \sum_{j = 1, j\neq i }^{N}{|\boldsymbol{h}^H_{i}\boldsymbol{v}_j|^2 }+ \sigma^2}\cdot
\end{equation}
We define $\boldsymbol{H} \triangleq [\boldsymbol{h}_1,\boldsymbol{h}_2,\ldots,\boldsymbol{h}_N]^T$, $\boldsymbol{V}~\triangleq [\boldsymbol{v}_1,\boldsymbol{v}_2,\ldots,\boldsymbol{v}_N]^T$, and $\boldsymbol{u}  \triangleq [u_1, u_2, \ldots,u_N]^T$. 



We seek to maximize the weighted sum rate (WSR) in the downlink channel subject to a total transmit power constraint, i.e., 
\begin{equation}\label{eq:SNR_maximization}
\begin{aligned}
\max_{\boldsymbol{V}} \quad & \sum_{i = 1}^{N}{\alpha_i\log_{2}{( 1 + \mathrm{SINR}_i)}} \\
\textrm{s.t.} \quad  & \mathtt{Tr}(\boldsymbol{V}\boldsymbol{V}^H)\leq P,
\end{aligned}
\end{equation}
where $\alpha_{i}$ indicates the user priority (assumed to be known) and where $P$ is the maximum transmit power at the base station. We assume the base station has perfect channel knowledge.
Problem~(\ref{eq:SNR_maximization}) is nonconvex and has been shown to be NP-hard~\cite{Liu2011,Luo2008}. 


\section{WMMSE algorithm }
The WMMSE algorithm addresses problem~(\ref{eq:SNR_maximization}) by solving a problem equivalent to it, i.e., with the same optimal $\boldsymbol{V}$, and amenable for block coordinate descent~\cite{Bertsekas99}.
The equivalent problem is given by a weighted sum mean square error minimization problem subject to a total transmit power constraint, i.e., 
\begin{subequations}\label{eq:sum-MSE_minimization}
\begin{alignat}{2}
\min_{\boldsymbol{u}, \boldsymbol{w}, \boldsymbol{V}} \quad & \sum_{i = 1}^{N}{\alpha_i \left(w_{i}e_{i}-\log_2{w_i} \right)}\\
\textrm{s.t.} \quad  & \textrm{Tr}(\boldsymbol{V}\boldsymbol{V}^H)\leq P,
\end{alignat}
\end{subequations}
where $e_i =\mathbb{E}_{\boldsymbol{x},n_i}\{{|\hat{x}_i-x_i|^2}\}$ assuming that $\boldsymbol{x}$ and $n_{i}$ are independent, where $\boldsymbol{x} \triangleq~[x_1,x_2,\ldots,x_N]^T$, where the estimated data symbol at the receiver is $\hat{x}_i = u_{i}y_{i} $, where $u_i \in \mathbb{C}$ is the receiver gain, where $w_i$ is the user weight, and where $\boldsymbol{w} \triangleq~[w_1,w_2,\ldots,w_N]^T$. 

Problem~(\ref{eq:sum-MSE_minimization}) is convex in each individual optimization variable, hence the partial optimization problem can be easily solved and a local optimum can be obtained by iteratively minimizing the cost function with respect to one variable while keeping the others fixed.

It results in the following sequential updates:
\begin{subequations}\label{eq:wmmse_update}
\begin{alignat}{3}
    {w}_i &= \frac{\sum_{j = 1}^{N}{|\boldsymbol{h}^H_i\boldsymbol{v}_j|^2} + \sigma^2}{\sum_{j = 1, j \neq i}^{N}{|\boldsymbol{h}^H_i\boldsymbol{v}_j|^2} + \sigma^2} \quad &\text{for}\ i = 1,\ldots,N,\label{eq:w_opt}\\
    {u}_i &= \frac{\boldsymbol{h}^H_i\boldsymbol{v}_i}{\sum_{j = 1}^{N}{|\boldsymbol{h}^H_i\boldsymbol{v}_j|^2} + \sigma^2} \quad &\text{for}\ i = 1,\ldots,N,\label{eq:u_opt}\\
    \boldsymbol{v}_i &= {\alpha_{i}u_{i}w_{i}\boldsymbol{h}_i}\left({ \boldsymbol{A} + \mu\boldsymbol{I}}\right)^{-1} \quad &\text{for}\ i = 1,\ldots,N,\label{eq:v_opt}
\end{alignat}
\end{subequations}
where 
\begin{equation}\label{eq:A}
\boldsymbol{A} \triangleq \sum_{i = 1}^{N}{\alpha_{i}w_{i}|u_i|^2\boldsymbol{h}_i\boldsymbol{h}^H_i}, 
\end{equation}
and where $\mu\geq0$ is a Langrange multiplier chosen such that the power constraint is satisfied. If $\mu= 0$ does not satisfy the power constraint, then the optimal $\boldsymbol{V}$ must satisfy the power constraint with equality. Hence, $\mu$ can be found by solving 
\begin{equation}
\mathtt{Tr}(\boldsymbol{V}\boldsymbol{V}^H)=P.
\end{equation}
As shown in~\cite{Shi2011}, this leads to equation
\begin{equation}\label{eq:eq_bisection}
\sum_{j = 1}^{M}{\frac{\mathtt{diag}(\boldsymbol{\Phi})_j}{\left( \mathtt{diag}(\boldsymbol{\Lambda})_{j} + \mu\right)^2}} = P,
\end{equation}
where $\mathtt{diag}(\boldsymbol{X})_{j}$ indicates the $j^{th}$ diagonal element of a matrix $\boldsymbol{X}$, where $\boldsymbol{U}\boldsymbol{\Lambda}\boldsymbol{U}^H$ is the eigendecomposition of $\boldsymbol{A}$, and where $\boldsymbol{\Phi} = \boldsymbol{U}^{H}(\sum_{i = 1}^{N}{\alpha^2_{i}w^2_{i}|u_i|^2\boldsymbol{h}_i\boldsymbol{h}_i^H})\boldsymbol{U}$.
The left hand-side of~(\ref{eq:eq_bisection}) is monotonically decreasing in $\mu$, therefore $\mu$ can be found by bisection search~\cite{rburden81} with the starting points
$$\mu_{\mathrm{low}} =\sqrt{\frac{1}{P}\sum_{j = 1}^{M}{\mathtt{diag}(\boldsymbol{\Phi}})_j} \quad \textrm{and}\quad \mu_{\mathrm{high}} = 0\,,$$ where $\mu_{\mathrm{low}}$ and $\mu_{\mathrm{high}}$ are such that the left-hand side of (\ref{eq:eq_bisection}) is smaller and greater than $P$, respectively.

To summarize, $\boldsymbol{V}$ is first initialized such that $\mathtt{Tr}(\boldsymbol{V}\boldsymbol{V}^{H})\leq~P$, then $\boldsymbol{w},\boldsymbol{u}$, and $\boldsymbol{V}$ are iteratively updated according to~(\ref{eq:wmmse_update}) until a convergence criterion is met.
For further details on the WMMSE algorithm, we refer the reader to~\cite{Shi2011}.

\section{Unfolded WMMSE algorithm}
 We propose  to unfold a fixed number of iterations of the WMMSE algorithm. 
The update equations of $\boldsymbol{w}$ and $\boldsymbol{u}$, i.e., (\ref{eq:w_opt}) and (\ref{eq:u_opt}), involve only simple vector and matrix operations, and can therefore be readily inserted in the unfolded computational graph. Conversely, the update equation of $\boldsymbol{V}$, i.e., (\ref{eq:v_opt}), involves a matrix inversion, an eigendecomposition, and a bisection search. It is complicated to unfold these operations~\cite{Sun2018}. Therefore, we propose to circumvent them by resorting to the projected gradient descent (PGD) approach~\cite{Bertsekas99}.

In the WMMSE algorithm, the step in (\ref{eq:v_opt}) is obtained as the solution to 
\begin{subequations}\label{eq:sum-MSE_minimization2}
\begin{alignat}{2}
\min_{\boldsymbol{V}} \quad & \sum_{i = 1}^{N}{\alpha_i \left(w_{i}e_{i}-\log_2{w_i} \right)}\\
\textrm{s.t.} \quad  & \textrm{Tr}(\boldsymbol{V}\boldsymbol{V}^H)\leq P,
\end{alignat}
\end{subequations}
with the method of Lagrange multipliers~\cite{Bertsekas1996}, where $\alpha_i$ is the user priority, where $w_i$ is the user weight, and where 
\begin{equation}
e_i =  \sum_{j=1}^{N}{|u_i\boldsymbol{h}_i^{H}\boldsymbol{v}_{j}|^2} -2 u_i \boldsymbol{h}_{i}^H\boldsymbol{v}_i +\sigma^2 |u_i|^2 +1.
\end{equation}
It leads to the Lagrangian function minimization, i.e.,
\begin{equation}\label{eq:lagrangian_for_v}
\begin{aligned}
\min_{\mu, \boldsymbol{V}} \quad & \sum_{i = 1}^{N}{\alpha_i \left(w_{i}e_{i}-\log_2{w_i} \right)} + \mu\mathtt{Tr}(\boldsymbol{V}\boldsymbol{V}^H) - \mu P,
\end{aligned}
\end{equation}
where $\mu$ is the Langrange multiplier. 
We observe in problem~(\ref{eq:sum-MSE_minimization2}) that i) the cost function is convex and differentiable and that ii) the constraint is a convex set. Therefore problem~(\ref{eq:sum-MSE_minimization2}) can be alternatively solved with the projected gradient descent approach. It is a first-order  method, thus it requires only gradient information and function values. At each iteration, the optimization variable is updated by i) taking a step in the descent direction defined by the negative gradient of the cost function and ii) projecting the
update onto the feasible set determined by the constraint.

We define $f(\boldsymbol{V}) \triangleq \sum_{i = 1}^{N}{\alpha_i \left(w_{i}e_{i}-\log_2{w_i} \right)}$ as our cost function and $\mathcal{C}= \{ \boldsymbol{V}| \mathtt{Tr}(\boldsymbol{V}\boldsymbol{V}^H)\leq P\}$ as the set defined by the power constraint. The $k^{th}$ PGD update is given by 
\begin{subequations}\label{eq:PGD}
\begin{alignat}{2}
\widetilde{\boldsymbol{V}}^{(k)} &= {\boldsymbol{V}}^{(k-1)} - {\gamma}\nabla f({{\boldsymbol{V}}}^{(k-1)}),\label{eq:PGD_first} \\
\boldsymbol{V}^{(k)} &= \Pi_{\mathcal{C}}\{\widetilde{\boldsymbol{V}}^{(k)}\}\label{eq:PGD_second},
\end{alignat}
\end{subequations}
where $\nabla f({{\boldsymbol{V}}}^{(k)})=~[\nabla f({{\boldsymbol{v}}}^{(k)}_1), \nabla f({{\boldsymbol{v}}}^{(k)}_2), \ldots,\nabla f({{\boldsymbol{v}}}^{(k)}_N)]^T $, where $ \nabla f({{\boldsymbol{v}}}^{(k)}_i) =-2\alpha_{i}w_{i}u_{i}\boldsymbol{h}_i + 2\boldsymbol{A}{\boldsymbol{v}}^{(k)}_i$, where $\boldsymbol{A}$ is defined in~(\ref{eq:A}), where ${\gamma}$ is the step size, and where ${\Pi_{\mathcal{C}}\{\boldsymbol{V}\}=\min_{\boldsymbol{Z} \in \mathcal{C}}{\left\lVert \boldsymbol{V} - \boldsymbol{Z} \right\rVert}}$. In particular,
\begin{equation}\label{eq:projection}
\Pi_{\mathcal{C}}\{\boldsymbol{V}\} = 
\begin{cases}
      \boldsymbol{V}, & \text{if}\ \mathtt{Tr}(\boldsymbol{V}\boldsymbol{V}^H)\leq P \\
      \frac{\boldsymbol{V}}{\left\lVert \boldsymbol{V} \right\rVert}\sqrt{P}, & \text{otherwise.}
    \end{cases} 
\end{equation}
As a result, we replace the matrix inversion, the eigendecomposition, and the bisection search required by~(\ref{eq:v_opt}) with simple vector operations, differentiable almost everywhere, that can be easily inserted in the unfolded computational graph.

To summarize, in this variant of the WMMSE, which we will refer to as \textit{unfolded WMMSE}, we first initialize $\boldsymbol{V}$  such that $\mathtt{Tr}(\boldsymbol{V}\boldsymbol{V}^{H})\leq~P$, then we sequentially compute  (\ref{eq:w_opt}), (\ref{eq:u_opt}), and $K$ PGD steps~(\ref{eq:PGD}) for a total of $L$ iterations. 
Consequently, a fixed number of operations is performed, namely a fixed number of iterations and of PGD steps per iteration. This yields to deterministic data flow, predetermined execution time, and fixed and known computational complexity.

\begin{algorithm}
\caption{Unfolded / Unfoldable WMMSE }
\hspace*{6pt} \textbf{Input: $\boldsymbol{H}$} 
\begin{algorithmic}
\STATE {Initialize $\boldsymbol{V}$ such that $\mathtt{Tr}(\boldsymbol{V}\boldsymbol{V}^{H})\leq P$}
\FOR{$l = 1,2,\dots,L$}
\FOR{$i = 1,2,\ldots,N$}
\STATE{${w}_i = \frac{\sum_{j = 1}^{N}{|\boldsymbol{h}^H_i\boldsymbol{v}_j|^2} + \sigma^2}{\sum_{j = 1, j \neq i}^{N}{|\boldsymbol{h}^H_i\boldsymbol{v}_j|^2} + \sigma^2}$}
\ENDFOR
\FOR{$i = 1,2,\ldots,N$}
\STATE{${u}_i= \frac{\boldsymbol{h}^H_i\boldsymbol{v}_i}{\sum_{j = 1}^{N}{|\boldsymbol{h}^H_i\boldsymbol{v}_j|^2} + \sigma^2}$}
\ENDFOR
\FOR{$k = 1,2,\ldots,K$}
\FOR{$i = 1,2,\ldots,N$}
\IF{$k == 1$}
\STATE{${\boldsymbol{v}_i}^{(0)} = {\boldsymbol{v}_i}$}
\ENDIF
\STATE{$\widetilde{\boldsymbol{v}_i}^{(k)} = {\boldsymbol{v}}_i^{(k-1)} - {\gamma}_{l}^{(k)}\nabla f({{\boldsymbol{v}}}_i^{(k-1)})$}
\ENDFOR
\STATE{$\boldsymbol{V}^{(k)} = \Pi_{\mathcal{C}}\{\widetilde{\boldsymbol{V}}^{(k)}\}$}
\ENDFOR
\STATE{${\boldsymbol{V}} = {\boldsymbol{V}^{(K)}}$}
\ENDFOR
\end{algorithmic}
\end{algorithm}

\section{Trainable unfolded WMMSE}
\begin{figure*}
         \centering
         \scalebox{0.29}{\input{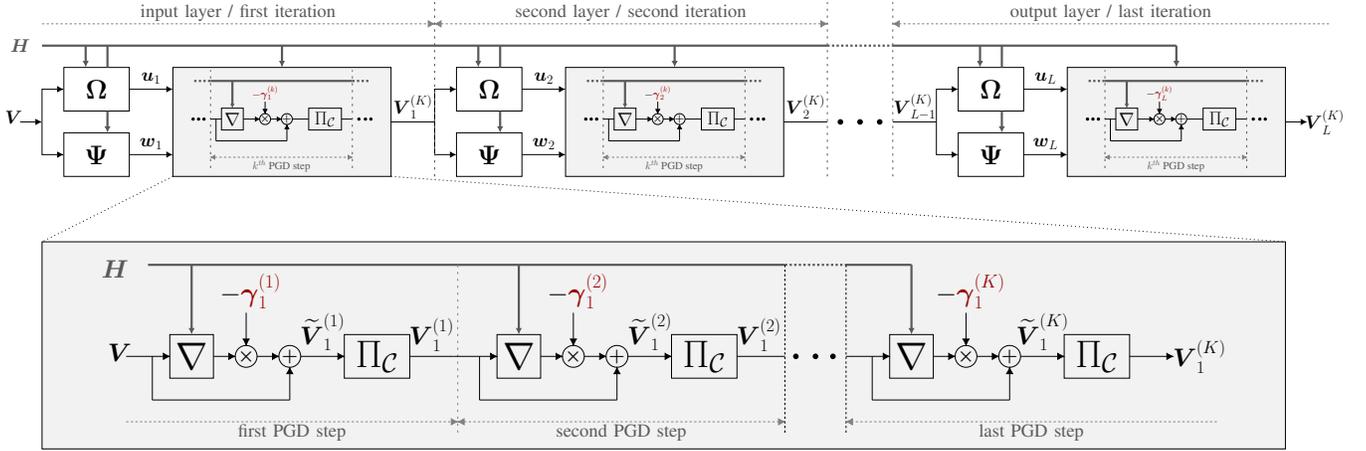}}
			\caption{The neural network architecture obtained by unfolding $L$ iterations of the WMMSE and $K$ projected gradient descent (PGD) steps per iteration. The subscript $(\cdot)_{l}$ corresponds to the $l^{th}$ layer/iteration and the superscript $(\cdot)^{(k)}$ corresponds to the $k^{th}$ PGD step. The trainable parameters are the step sizes of the PGD approach. Each layer of the neural network is given by the update equation of $\boldsymbol{w}$~(\ref{eq:w_opt}), indicated by $\boldsymbol{\Psi}$, by the update equation of $\boldsymbol{u}$~(\ref{eq:u_opt}), indicated by $\boldsymbol{\Omega}$, and by unfolding $K$ PGD steps, as depicted in the gray box. In particular, $\nabla$ and $\Pi_\mathcal{C}$ indicate the gradient and the projection operations in~(\ref{eq:PGD}), respectively.   }
			\vspace{-1.8mm}
         	\label{fig:architecture}        	
\end{figure*} 
It is clear that the performance of the unfolded WMMSE is penalized by the computational constraint that we impose when setting the number of iterations and the number of PGD steps per iteration. Therefore, the original WMMSE algorithm constitutes an upper bound on the performance of the unfolded WMMSE. However, how far from the upper bound the actual performance is depends on the choice of the step sizes used in the projected gradient descent method. Considering that we truncate the sequence of PGD steps to $K$, the role of the step sizes is even more relevant, as they regulate the behavior and the convergence speed of the sequence of PGD steps. We propose to optimize the performance of the unfolded WMMSE by means of deep learning, i.e., by making the step sizes trainable parameters of a deep learning architecture. We collectively denote them as $\boldsymbol{\Gamma}=~[\boldsymbol{\gamma}_{1}, \boldsymbol{\gamma}_{2},\ldots,\boldsymbol{\gamma}_{L}]$, where $\boldsymbol{\gamma}_{l} = [\gamma^{(1)}_{l},\gamma^{(2)}_{l},\ldots,\gamma^{(K)}_{l}]$ and where $\gamma^{(k)}_{l}$ is the step size used by the $k^{th}$ PGD step in the $l^{th}$ iteration.  Algorithm 1 reports the pseudocode of the unfoldable WMMSE and the corresponding computational graph defines a deep neural network. Each iteration of the unfolded WMMSE corresponds to a layer of the neural network. Thus, each layer consists of the update equation of $\boldsymbol{w}$~(\ref{eq:w_opt}), the update equation of $\boldsymbol{u}$~(\ref{eq:u_opt}), and $K$ unfolded PGD steps. Fig.~\ref{fig:architecture} depicts the overall neural network architecture.

Although the projected gradient descent is used as an alternative approach to solve problem~(\ref{eq:sum-MSE_minimization2}), which is optimally solved by~(\ref{eq:v_opt}), we do no train $\boldsymbol{\Gamma}$ such that the truncated PGD sequence  can  give as close an approximation of~(\ref{eq:v_opt}) as possible. We train $\boldsymbol{\Gamma}$ such that the unfolded WMMSE can achieve the highest possible WSR within the fixed network architecture and computational complexity. The training is necessarily unsupervised, as the optimal transmit beamformer is unknown.
Therefore, in order to learn $\boldsymbol{\Gamma}$, the natural choice of loss function is the following, i.e., 
\begin{equation}\label{eq:loss}
\mathcal{L}(\boldsymbol{\Gamma}) = -\frac{1}{N_{\mathrm{s}}} \sum_{n= 1}^{N_{\mathrm{s}}}{\sum_{l = 1}^{L}{f_{\mathrm{WSR}}(\boldsymbol{H}_{n}},\boldsymbol{V}_{l}};\boldsymbol{\Gamma}),   
\end{equation}
where $N_{\mathrm{s}}$ is the size of the training set and where $f_{\mathrm{WSR}}(\boldsymbol{H}_{n},\boldsymbol{V}_{l};\boldsymbol{\Gamma})$ indicates the WSR~(\ref{eq:SNR_maximization}) achieved with the $n^{th}$ channel realization, drawn i.i.d.\ from the distribution of $\boldsymbol{H}$, and with the transmit beamformer given as output by the $l^{th}$ layer of the neural network, hence the parameterization with $\boldsymbol{\Gamma}$. As can be noted, we use a loss function that takes into account the transmit beamformer given at each layer. In this way, as in~\cite{Wiesel2019}, we mitigate possible complications that occur when training, such as vanishing gradients and saturation of hidden units~\cite{Glorot10}. It must also be noted that we do not impose any constraint on the step sizes across the layers. Each of them is allowed to assume any value in $\mathbb{R}$ while minimizing $\mathcal{L}(\boldsymbol{\Gamma})$. The loss function $\mathcal{L}(\boldsymbol{\Gamma})$ is continuous and differentiable almost everywhere with respect to $\boldsymbol{\Gamma}$, hence we can use gradient-based
optimization methods~\cite{Gregor2010} for training. In particular, we minimize $\mathcal{L}(\boldsymbol{\Gamma})$ by applying a variant of the stochastic gradient descent method, i.e., the Adam optimizer~\cite{adam2014}.

\section{Numerical results}
\subsection{Setup}
The proposed unfolded WMMSE algorithm, illustrated by the network structure in Fig.~\ref{fig:architecture}, was implemented in Python 3.6.8 with Tensorflow 1.13.1~\cite{tensorflow2015-whitepaper} and the original WMMSE algorithm was implemented in Python 3.6.8 as well. The full code is available at~\cite{GITHUB}.
In the simulation settings, we consider four transmit antennas at the base station and four single-antenna users, i.e., $M=4$ and $N=4$, respectively, and we set the user priorities to one, i.e., $\alpha_i = 1$ for $i=1,\ldots,N$.
As in~\cite{Huang2019,Brandt2015,Christensen2008}, we initialize the WMMSE and the unfolded WMMSE algorithms with matched filtering, i.e., $\boldsymbol{V}=a\boldsymbol{H}$, where $a \in \mathbb{R}$ is chosen such that the power constraint is satisfied with equality.
In the WMMSE algorithm, we end the bisection search used to solve~(\ref{eq:eq_bisection}) when the equality constraint is satisfied up to an error of $10^{-4}$ and, unless otherwise stated, we stop the iterations when the increment in WSR at the next iteration is less than or equal to $10^{-4}$. It must be noted that the unfolded WMMSE algorithm can be readily implemented by resorting to basic neural network operations. The update equations of $\boldsymbol{w}$ (\ref{eq:w_opt}) and $\boldsymbol{u}$ (\ref{eq:u_opt}) involve only vector multiplications and summations and the update equations of $\boldsymbol{V}$ involve multiplications and summations (\ref{eq:PGD_first}) and a projection~(\ref{eq:PGD_second}), which can be equivalently implemented with elemental neural network operations, without employing conditional statements, i.e.,  
$$
\Pi_{\mathcal{C}}\{\boldsymbol{V}\}= \frac{\boldsymbol{V}\sqrt{P}}{\phi({\left\lVert \boldsymbol{V} \right\rVert}-\sqrt{P})+\sqrt{P}}\raisepunct{,}
$$
where $\phi(x) = \max(0,x)$ denotes the rectified linear unit (ReLU).
As available deep learning tools do not support complex numbers, an alternative representation to map the complex variables and operations of the unfolded WMMSE algorithm to the real domain must be adopted. To this end, we apply the decomposition given by $u = [\Re(u), \ \Im(u)]$, $w = [\Re(w), \ \Im(w)]$, $\boldsymbol{v} = [\Re(\boldsymbol{v})^T, \ \Im(\boldsymbol{v})^T]^T$, and 
$$
\boldsymbol{h} = \begin{bmatrix} \Re(\boldsymbol{h}) & -\Im(\boldsymbol{h}) \\ \Im(\boldsymbol{h}) & \Re(\boldsymbol{h}) \end{bmatrix}.
$$
Unless otherwise stated, we initialize the trainable step sizes to one, i.e., $\gamma^{(k)}_l = 1$ for $k=1,\ldots,K$ and $l=1,\ldots,L$. We set the learning rate of the Adam optimizer to $10^{-3}$. The training set and the test set consist of channel realizations drawn i.i.d.\ from a complex standard  Gaussian distribution.
We consider up to six iterations, i.e., $L = 1,\ldots,6$, four to eight PGD steps, i.e., $K = 4,\ldots,8$, and seven different signal-to-noise ratio (SNR) values equally spaced from $5$ dB to $20$ dB. We train a different unfolded neural network for each considered combination. 
The size of the training set varies from $2\cdot10^6$ to $8\cdot10^6$ channel realizations depending on the considered values of $L$, $K$, and SNR. As the values of $L$, $K$, and SNR increase, we use a larger training set. One training batch consists of 100 channel realizations. 
The presented results are averaged on a test set of $10^5$ channel realizations.

\subsection{Results and Discussion}
We assess the performance of the proposed unfolded WMMSE algorithm as the number of unfolded iterations, i.e., of layers, varies and we compare its performance with i) the original WMMSE algorithm truncated at the same number of iterations and with ii) the unfolded WMMSE algorithm with the trainable step sizes constrained to be equal across all PGD steps of the same layer, i.e., $\gamma^{(1)}_{l}=\ldots=\gamma^{(K)}_{l} \text{for}\ l= 1,\ldots,L$.
\begin{figure}[ht]
         \centering
         	\scalebox{1.0}{
\begin{tikzpicture}

\begin{axis}[
legend cell align={left},
legend style={fill opacity=0.8, draw opacity=1, text opacity=1, at={(0.99,0.01)}, anchor=south east, draw=black},
tick align=outside,
tick pos=left,
x grid style={white!69.0196078431373!black},
xlabel={Number of iterations $L$},
xmajorgrids,
xmin=-0.25, xmax=5.25,
xtick style={color=black},
xtick={0,1,2,3,4,5},
xticklabels={1,2,3,4,5,6},
y grid style={white!69.0196078431373!black},
ylabel={Weighted sum rate [bit/s]},
ymajorgrids,
ymin=7.47429355197759, ymax=9.97809109195418,
ytick style={color=black}
]

\addplot [thick, blue, mark=*, mark size=3, mark options={solid}]
table {%
0 8.55238878736002
1 9.31791222232729
2 9.54739862646093
3 9.65311001630277
4 9.71138830671989
5 9.74597729840455
};
\addlegendentry{\scriptsize Unfolded WMMSE}

\addplot [thick, red, mark=triangle*, mark size=3, mark options={solid}]
table {%
0 7.94556340991089
1 9.10786237131682
2 9.48402979482838
3 9.63048534031205
4 9.70503811522628
5 9.74962895835738
};
\addlegendentry{\scriptsize WMMSE}

\addplot [thick, green!50.1960784313725!black, mark=square*, mark size=2.5, mark options={solid}]
table {%
0 7.58810253106744
1 8.6926536988122
2 9.15003443838656
3 9.37242139244959
4 9.49881819729442
5 9.57269861820653
};
\addlegendentry{ \scriptsize Unfolded WMMSE - same $\gamma$}

\addplot [very thick, black, dashed]
table {%
0 9.86428211286433
1 9.86428211286433
2 9.86428211286433
3 9.86428211286433
4 9.86428211286433
5 9.86428211286433
};
\addlegendentry{\scriptsize WMMSE at convergence}
\end{axis}

\end{tikzpicture}}
			\caption{Weighted sum rate obtained with $N = 4$, $M = 4$, SNR of 10 dB, and $K=4$. The label `same $\gamma$' indicates that the same step size is used across all the PGD steps of the same layer, but the step sizes are different across the layers. }
         	\label{fig:WSR_10db}        	
\end{figure}
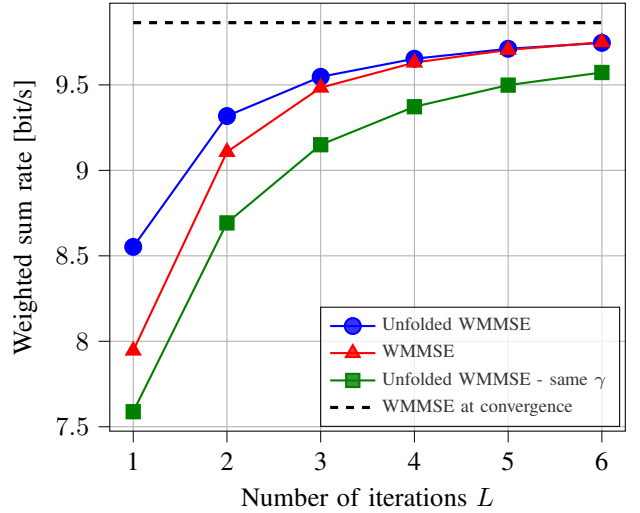

\begin{figure}[ht]
         \centering
         	\scalebox{1}{\input{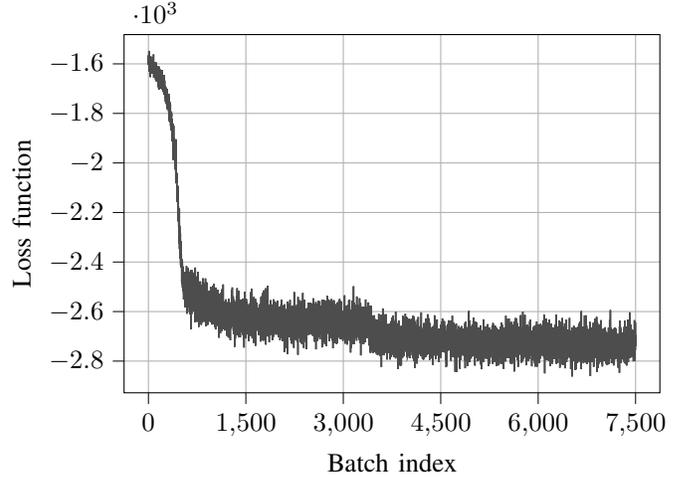}}
			\caption{Training loss function of the proposed unfolded WMMSE algorithm, with $N = 4$, $M = 4$, SNR of 10 dB, $L=3$, and $K=4$.}
         	\label{fig:training_loss_7500}        	
\end{figure}
Fig.~\ref{fig:WSR_10db} shows the performance of the different approaches as the number of iterations varies, with SNR of 10 dB and with $K = 4$, i.e., with 4 PGD steps in each layer of the unfolded WMMSE algorithm. The WSR attained by the WMMSE at convergence, which constitutes an upper bound for the proposed unfolded WMMSE, is reported for reference. As can be seen, the unfolded WMMSE surpasses the performance of the WMMSE up to $L=4$ and performs equally  to it for $L=5$ and for $L=6$. It can also be noted that for $L=6$ both the WMMSE and the unfolded WMMSE attain the 98 percent of the WSR achieved at convergence.
Furthermore, the benefit of imposing no constraints on the trained step sizes can be observed as the unfolded WMMSE with constrained step sizes fails to surpass the WMMSE or to perform comparably to it.
It naturally follows that, if the constraint is extended to the layers, i.e., the step sizes are restricted to be equal across the PGD steps and across the layers, the attained WSR would be even lower. Fig.~\ref{fig:training_loss_7500} shows the loss function $\mathcal{L}(\boldsymbol{\Gamma})$ when training the proposed unfolded WMMSE algorithm, with $L=3$.
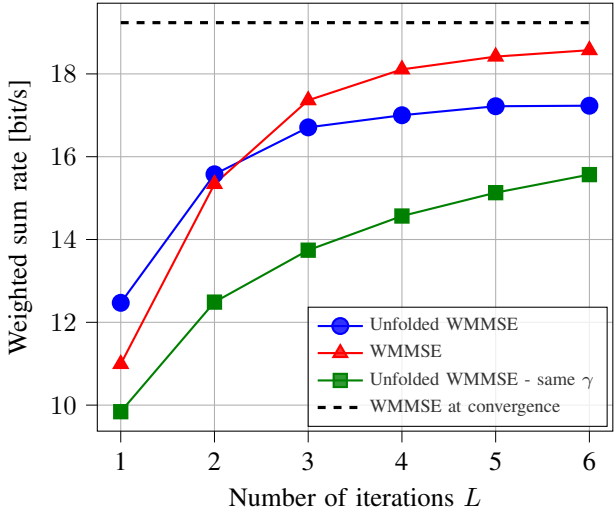
\begin{figure}[t]
         \centering
         	\scalebox{1.0}{
\begin{tikzpicture}

\begin{axis}[
legend cell align={left},
legend style={fill opacity=0.8, draw opacity=1, text opacity=1, at={(0.99,0.01)}, anchor=south east, draw=black},
tick align=outside,
tick pos=left,
x grid style={white!69.0196078431373!black},
xlabel={Number of iterations $L$},
xmajorgrids,
xmin=-0.25, xmax=5.25,
xtick style={color=black},
xtick={0,1,2,3,4,5},
xticklabels={1,2,3,4,5,6},
y grid style={white!69.0196078431373!black},
ylabel={Weighted sum rate [bit/s]},
ymajorgrids,
ymin=9.37117582040136, ymax=19.7075466749937,
ytick style={color=black}
]

\addplot [thick, blue, mark=*, mark size=3, mark options={solid}]
table {%
0 12.4708852016052
1 15.5750042200248
2 16.7086710437917
3 17.0015535957825
4 17.2190194449553
5 17.2314861828137
};
\addlegendentry{\scriptsize Unfolded WMMSE}

\addplot [thick, red, mark=triangle*, mark size=3, mark options={solid}]
table {%
0 10.9921967126972
1 15.3398891414424
2 17.3626205644617
3 18.1085861095798
4 18.4175821599519
5 18.57297621584
};
\addlegendentry{\scriptsize WMMSE}

\addplot [thick, green!50.1960784313725!black, mark=square*, mark size=2.5, mark options={solid}]
table {%
0 9.84101085924647
1 12.4878643449648
2 13.7419801311465
3 14.5670963987776
4 15.1311066425139
5 15.5676164705176
};
\addlegendentry{\scriptsize Unfolded WMMSE - same $\gamma$}

\addplot [very thick, black, dashed]
table {%
0 19.2377116361486
1 19.2377116361486
2 19.2377116361486
3 19.2377116361486
4 19.2377116361486
5 19.2377116361486
};
\addlegendentry{\scriptsize WMMSE at convergence}

\end{axis}

\end{tikzpicture}}
			\caption{Weighted sum rate obtained with $N = 4$, $M = 4$, SNR of 20 dB, and $K=4$. The label `same $\gamma$' indicates that the same step size is used across all the PGD steps of the same layer, but the step sizes are different across the layers. }
         	\label{fig:WSR_20db}        	
\end{figure}

Fig.~\ref{fig:WSR_20db} shows the same comparison as Fig.~\ref{fig:WSR_10db}, but with a higher SNR, i.e., of 20 dB. As before, the unfolded WMMSE with constrained step sizes is outperformed by the WMMSE for any value of $L$. For $L=1$ and $L=2$ the proposed unfolded WMMSE obtains an improved WSR over the WMMSE, but as the number of iterations grows, the WMMSE surpasses the proposed unfolded WMMSE. It indicates that, at higher SNR, solving the inner optimization problem~(\ref{eq:sum-MSE_minimization2}) to a higher precision by resorting to the method of Lagrange multipliers, as in the original WMMSE algorithm, has benefits. However, at a fixed computational complexity, a larger number of iterations of the unfolded WMMSE can be executed with respect to the original WMMSE. In fact, the projected gradient descent in the unfolded WMMSE involves only simple vector operations~(see (\ref{eq:PGD})), while the method of Lagrange multipliers requires the complex matrix inversion and eigendecomposition~(see (\ref{eq:v_opt})).

Looking at complexity in terms of big O notation, the unfolded WMMSE algorithm has complexity of $\mathcal{O}(LKM^2)$, while the original WMMSE algorithm has complexity of $\mathcal{O}(LM^3)$. However, big O notation hides the fact that the operations in the unfolded WMMSE algorithm are internally non-iterative in nature and thus suitable for implementation on standardized hardware optimized for deep learning processing. Conversely, the operations used in the original WMMSE algorithm, such as the eigendecomposition, have complexity of $\mathcal{O}(M^3)$ but are typically implemented using iterative procedures.

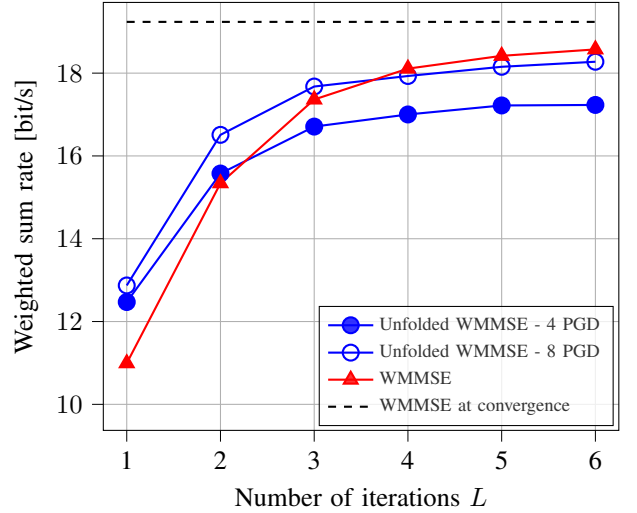
\begin{figure}[ht]
         \centering
         	\scalebox{1.0}{
\begin{tikzpicture}

\begin{axis}[
legend cell align={left},
legend style={fill opacity=0.8, draw opacity=1, text opacity=1, at={(0.99,0.01)}, anchor=south east, draw=black},
tick align=outside,
tick pos=left,
x grid style={white!69.0196078431373!black},
xlabel={Number of iterations $L$},
xmajorgrids,
xmin=-0.25, xmax=5.25,
xtick style={color=black},
xtick={0,1,2,3,4,5},
xticklabels={1,2,3,4,5,6},
y grid style={white!69.0196078431373!black},
ylabel={Weighted sum rate [bit/s]},
ymajorgrids,
ymin=9.37117582040136, ymax=19.7075466749937,
ytick style={color=black}
]
\addplot [thick, blue, mark=*, mark size=3, mark options={solid}]
table {%
0 12.4708852016052
1 15.5750042200248
2 16.7086710437917
3 17.0015535957825
4 17.2190194449553
5 17.2314861828137
};
\addlegendentry{\scriptsize Unfolded WMMSE - 4 PGD}
\addplot [thick, blue, mark=*, mark size=3, mark options={solid,fill opacity=0}]
table {%
0 12.8716090978318
1 16.5091810381069
2 17.6763175681697
3 17.9306262847849
4 18.1530108179765
5 18.2748562726735
};
\addlegendentry{\scriptsize Unfolded WMMSE - 8 PGD}
\addplot [thick, red, mark=triangle*, mark size=3, mark options={solid}]
table {%
0 10.9921967126972
1 15.3398891414424
2 17.3626205644617
3 18.1085861095798
4 18.4175821599519
5 18.57297621584
};
\addlegendentry{\scriptsize WMMSE}
\addplot [thick, black, dashed]
table {%
0 19.2377116361486
1 19.2377116361486
2 19.2377116361486
3 19.2377116361486
4 19.2377116361486
5 19.2377116361486
};
\addlegendentry{\scriptsize WMMSE at convergence}
\end{axis}

\end{tikzpicture}}
			\caption{Weighted sum rate obtained with $N = 4$, $M = 4$, and SNR of 20~dB.}
     	\label{fig:WSR_20db_PGD}        	
\end{figure}

We can improve the solution to~(\ref{eq:sum-MSE_minimization2}) in the unfolded WMMSE algorithm by extending the sequence of PGD steps. However, as the number of PGD steps included in the optimization grows, we empirically observe that the neural network converges with difficulty to a good optimum~\cite{Glorot10}. A feasible approach to address this complication consists of progressively adding a single step size and jointly training the newly added step size with the pretrained step sizes, initialized to their optimized values. Fig.~\ref{fig:WSR_20db_PGD} shows the substantial improvement in WSR attained by progressively extending the initial PGD sequence up to eight steps and by training the added step sizes as described. The unfolded WMMSE surpasses the original WMMSE up to $L=3$ and for higher values of $L$ reaches an almost comparable WSR.

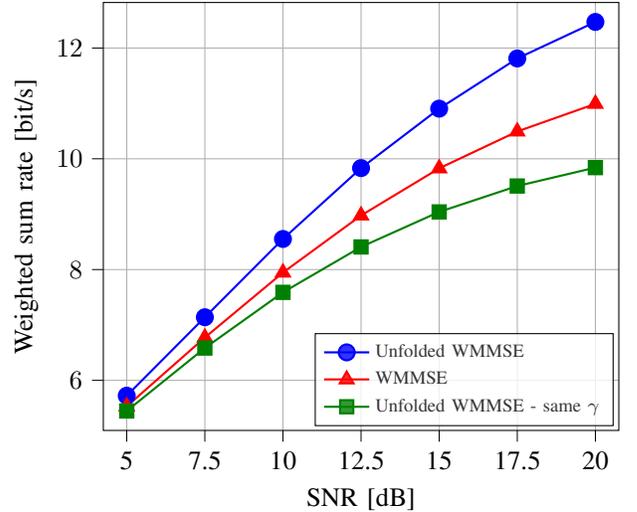
\begin{figure}[ht]
         \centering
         	\scalebox{1.0}{
\begin{tikzpicture}

\begin{axis}[
legend cell align={left},
legend style={fill opacity=0.8, draw opacity=1, text opacity=1, at={(0.99,0.01)}, anchor=south east, draw=black},
tick align=outside,
tick pos=left,
x grid style={white!69.0196078431373!black},
xlabel={SNR [dB]},
xmajorgrids,
xmin=-0.3, xmax=6.3,
xtick style={color=black},
xtick={0,1,2,3,4,5,6},
xticklabels={5,7.5,10,12.5,15,17.5,20},
y grid style={white!69.0196078431373!black},
ylabel={Weighted sum rate [bit/s]},
ymajorgrids,
ymin=5.09527954976594, ymax=12.8221045183595,
ytick style={color=black}
]
\addplot [thick, blue, mark=*, mark size=3, mark options={solid}]
table {%
0 5.72214184840159
1 7.13891467534541
2 8.55238878736002
3 9.83180849371341
4 10.9054756790702
5 11.8118736693653
6 12.4708852016052
};
\addlegendentry{\scriptsize Unfolded WMMSE}

\addplot [thick, red, mark=triangle*, mark size=3, mark options={solid}]
table {%
0 5.52920786702354
1 6.77231149599143
2 7.94556340991089
3 8.97640887611549
4 9.82715410871368
5 10.493111606781
6 10.9921967126972
};
\addlegendentry{\scriptsize WMMSE}

\addplot [thick, green!50.1960784313725!black, mark=square*, mark size=2.5, mark options={solid}]
table {%
0 5.44649886652019
1 6.5828710301115
2 7.58810253106744
3 8.40839215193566
4 9.04086183774931
5 9.50829837466094
6 9.84101085924647
};
\addlegendentry{\scriptsize Unfolded WMMSE - same $\gamma$}

\end{axis}

\end{tikzpicture}}
			\caption{Weighted sum rate obtained with $N = 4$, $M = 4$, $L=1$, and $K=4$. The label `same $\gamma$' indicates that the same step size is used across all the PGD steps of the same layer.}
         	\label{fig:one_outer_loop_no_ref}        	
\end{figure}

Finally, Fig.~\ref{fig:one_outer_loop_no_ref} shows the WSR attained under different SNR conditions, with $L=1$ and $K=4$. As expected, the WSR increases as the SNR increases. Also, the trend previously observed is confirmed as, for all SNR values, i) the  unfolded WMMSE with  constrained  step  sizes  is  outperformed  by the original WMMSE and ii) the proposed unfolded WMMSE surpasses the original WMMSE with a performance gap that increases with the SNR.

\section{Conclusion}
We provided a formulation of the WMMSE beamforming algorithm that allows for the novel application of deep unfolding, with the aim to address the complexity versus performance trade-off at the base station. 
Simulation results confirmed that deep unfolding can successfully address this trade-off, by leveraging expert knowledge and incorporating it in the learning process and in the structure of the unfolded neural network. This interplay between deep learning and domain knowledge has multiple benefits with respect to end-to-end learning solutions already proposed in the literature as lower-complexity approximators of the WMMSE algorithm or as comparably performing alternatives. Moreover, the unfolded network is suitable for implementation on standardized hardware optimized for deep learning processing, envisioned to appear in the base stations in the near future. 

We focused on a single-base-station scenario, but our work can be easily extended to the multi-base-station case by changing the power constraint accordingly. 
Interesting research directions include i) the extension of our approach to the multiple-input multiple-output (MIMO) case and ii) the investigation of first order methods alternative to the projected gradient descent to solve the inner convex optimization in the original WMMSE algorithm.

\bibliographystyle{IEEEtran}
\bibliography{wmmse.bib}

\end{document}